\documentclass{amsart}
\newenvironment{nouppercase}{%
  \renewcommand{\uppercasenonmath}[1]{}}{}
\usepackage{enumerate,mathrsfs}
\usepackage{braket}
\usepackage{color}
\usepackage{graphicx}
\usepackage{tikz}
\usetikzlibrary{calc, shapes, positioning,backgrounds,arrows,automata,backgrounds,fit,decorations.pathreplacing}
\renewcommand{\subset}{\subseteq}

\DeclareMathOperator{\Aut}{Aut}

\DeclareMathOperator{\spn}{span}

\theoremstyle{definition}
\newtheorem{defn}{Definition}
\newtheorem{example}{Example}
\theoremstyle{theorem}
\newtheorem{theorem}{Theorem}

\title{\text{Quantum walks on regular graphs with realizations in a system of anyons}}
\author{Radhakrishnan Balu$^{\dag}$}
\address{$^{\dag}$Computer and Information Sciences Directorate, Army Research Laboratory,
Adelphi, MD, 21005-5069, USA.}
\email{radhakrishnan.balu.civ@mail.mil}
\address{$^{\dag}$Department of Mathematics, University of Maryland, College Park, MD 20742}
\email{rbalu@math.umd.edu}

\begin{document}
\begin{nouppercase}
\maketitle
\end{nouppercase}
\begin{abstract}
We build interacting Fock spaces from association schemes and set up quantum walks on the resulting regular graphs (distance-regular and distance-transitive). The construction is valid for growing graphs and the interacting Fock space is well defined asymptotically for the growing graph. To realize the quantum walks defined on the spaces in  terms of anyons we switch to the dual view of the association schemes and identify the corresponding modular tensor categories from the Bose-Mesner algebra. Informally, the fusion ring induced by the association scheme and a topological twist can be the basis for developing a modular tensor category and thus a system of anyons. Finally, we demonstrate the framework in the case of Grover quantum walk on distance-regular graph in terms of anyon systems for the graphs considered. In the dual perspective interacting Fock spaces gather a new meaning in terms of anyon collisions for the case of distance-regular graphs.
\end{abstract}
\section {Introduction}
In this work our contribution is to construct interacting Fock spaces (IFS) out of association schemes and for the particular case of IFS that correspond to distance-regular graphs we identify fusion rings by exploiting the duality of the association schemes. As an application of this framework we demonstrate the usual quantum walks defined on distance-regular graphs recast as IFS evolutions in terms of anyon systems of topological computation. We briefly recollect important aspects of quantum walks and topological computation before setting up the mathematical framework. After introducing the quantum walk example we offer few comments on the connection between distance-regular graphs and Temperley-Lieb algebras so as to make the connection with topological computation natural. There are several excellent sources on introducing subfactors and Temperley-Lieb algebras and we refer to Bisch \cite {Bisch2002} as a starting point. Anyon based topological computation can be described at several levels from cobordism based topological quantum field theory, subfactors of von Nuemann algebras, and tensor modular categories (TMC) with the later being the most studied formalism. There is more to TMCs than fusion rules and topological twists and we only focused here on these two parts as another way to approach TMCs. 

Quantum walks are important tools in deriving simulation protocols, perfect state transfer, and they also provide a framework for universal computation. Quantum walks involve a coin Hilbert space and a walker{'}s Hilbert space where the dynamics happens \cite{Salvador2012}. Let $\mathscr{C}^2$ (complex space) and $\mathbb{Z}$ (set of integers) correspond to Hilbert spaces of the coin and walker respectively. The dynamics of the quantum walk is described by the unitary operator U
\begin {align} \label {QWalk}
L^{\pm}(x) &= x\pm{1}, x\in{\mathbb{Z}}. \\
\mathscr{C}^2 &= \Pi_0\oplus{\Pi_1},\Pi_i = \ket{i}\bra{i}. \\
S(x) &= \Pi_0\otimes{L^+}+\Pi_1\otimes{L^-}. \\
U(x) &= S(x)T.
\end {align} 
That is, the walker takes a step to the right on $\mathbb{Z}$ if the outcome of tossing the quantum coin 
\begin {align*}
T &= \begin {bmatrix} a & -b^* \\
                                 b &  a^* \\
                  \end {bmatrix}, \\
|a|^2 + |b|^2 &= 1                  
\end {align*}
on $\mathscr{C}^2$ is 0 and moves a step left otherwise. A split step quantum walk that will later be used to derive the Dirac equation is defined as below:
\begin {align} \label {SSQWalk}
S(x) &= \Pi_0^{\theta_1}\otimes{L^+}+\Pi_1^{\theta_1}\otimes\mathbb{I}+\Pi_0^{\theta_2}\otimes\mathbb{I}+\Pi_1^{\theta_2}\otimes{L^-}. \\
U(x) &= S(x)T(\theta_2)S(x)T(\theta_1).
\end {align} 

\begin{figure}
\includegraphics[width=\columnwidth]{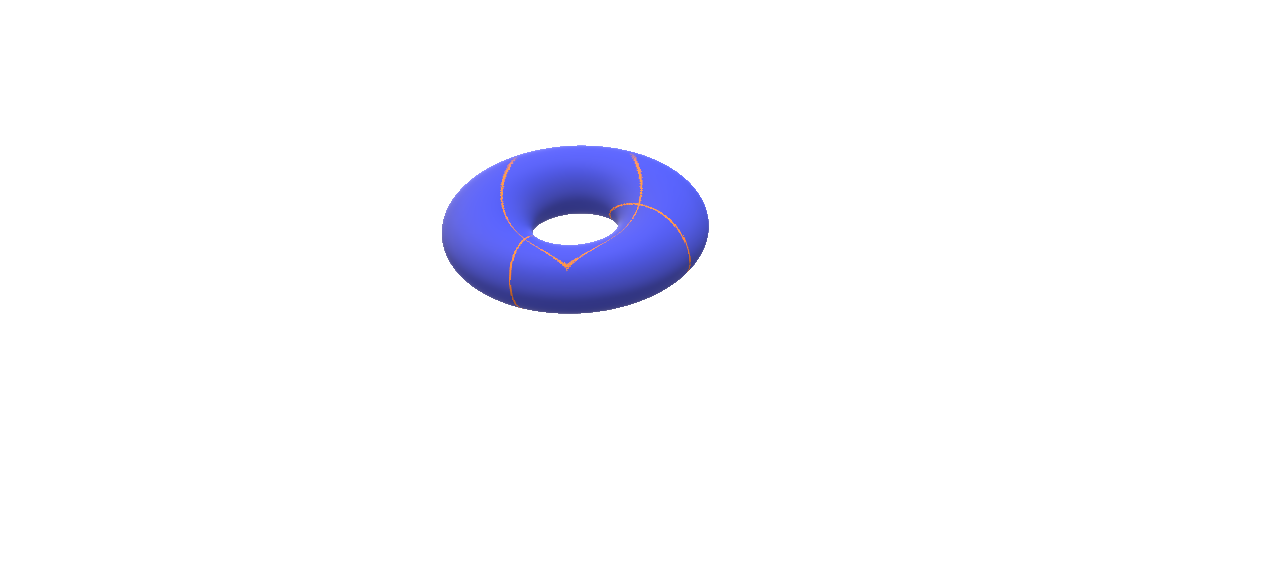}
\caption{ \label{fig:WLoop} Wilson loops of anyonic quasiparticles evolving on 2-Torus surface with one loop in the axial direction and the other perpendicular to it.}
\end{figure}

Topological quantum computation (TQC) is a special form of microscopic information processing that is robust against local perturbations.It has a steeper learning curve than circuits based quantum computation. To mitigate this and develop a better intuition for the computation, we can look at the dual process described on a graph. The physics behind this information process involves quantum particles that conform to statistics between that of bosons and that of fermions. Microscopically identical particles when permuted with respect to their positions, create statistical ensembles that are dependent on the type of particle. For example, in three dimensions, when two bosons are switched in configuration, the resulting wavefunction of the pair is the same as that of the original. On the other hand, fermionic wavefunctions acquire a phase of $\pi$ when switched similarly. Anyons are even stranger since can be antiparticles to themselves, and can produce any phase between zero and $\pi$ when evolved on a two dimensional surface and thus the name. From this simple description it is clear that the initial and final positions of the particles are important for the phase.The exact path traversed by them is irrelevant, hence topological, in situations where we want to use them for information processing. To have a path independent evolution we need a non contractible surface such as the 2-Torus instead of 2-Sphere in dimension two so as to have homotopically inequivalent paths. A pair of anyons when they collide with each other, may form another type of anyon (fusion) or a complimentary splitting event (graphically shaped like a pair-of-pants) from a single particle can happen with probabilities attached to the events. Topological quantum computation is organized with a finite set of anyons of different types and allowed to evolve, fuse together, and ending with a measurement forming what are called Wilson loops (Figure \ref{fig:WLoop}). Typically, the process starts with the production of a particle and antiparticle pair (quasiparticles to be precise), both are allowed to evolve along arbitrary paths which is equivalent to an unitary quantum gate (as the quantum evolution is a closed system) that encode algorithms, and finally fused together followed by measurements. It is a bit difficult to follow through the combinatorial events of anyon fusions, braidings (switching positions), and splitting in addition to keep track of the statistics that form a topological quantum computation. We describe the dual process on graphs generated by association schemes to develop another perspective and intuition on TQC. The fusion and splitting events form the nodes of the graphs that can support quantum probability spaces enabling the asymptotic analysis of TQC. One of the well studied anyon systems, the Ising model, consists of two particles apart from the vacuum 1 has the following fusion rules of the majorana fermion and the non-abelian anyon $\psi, \sigma, 1$:
\begin {align*}
\sigma \times \sigma &= 1 + \psi. \\
\sigma \times \psi &= \sigma. \\
\psi \times \psi &= 1. 
\end {align*}
The computational model is complete with the specification the S matrix (the columns of which are simultaneous eigen vectors of commuting fusion rules) and the $\theta$ twist (topological spin):
\begin {align*}
S &= \begin{bmatrix} 1 & \sqrt{2} & 1 \\ \sqrt{2} & 0 & -\sqrt{2} \\ 1 & -\sqrt{2} & 1 \end {bmatrix}.  \\
\theta_\sigma &= e^{\frac{i\pi} {8}} . \\
\theta_\psi &= -1.
\end {align*}
The fusion rules can also be expressed in terms of integers N (later interpreted as Krein parameters) as:  
\begin {align}  \label {FusionRules}
N^1_{\sigma \sigma} &= 1. \\  \nonumber
N^{\psi}_{\sigma \sigma} &= 1. \\ \nonumber
N^{\sigma}_{\sigma 1} &= 1. \\  \nonumber
N^{\sigma}_{\sigma \psi} &= 1. \\ \nonumber
\end {align}
This results in the dimension of the anyon $d_\sigma d_\sigma = \sum_c N^c_{ab} d_c$ as $d_\sigma = \sqrt{2}$ and for a quick introduction to topological computation the work of Rowell \cite {Rowell2012} is a good place to start.

To describe the topological process in algebraic terms we can consider an association scheme \cite {Radbalu2020} which is either a collection of adjacency matrices of graphs with a common set of $|\mathfrak{X}| = d$ vertices or it encodes 1-distance, 2-distance, ..., d-distance matrices of the same graph.
Let $X$ be a (finite) vertex set, and let $\mathfrak{X} = \{A_j\}_{j=0}^d$ be a collection of $X\times X$ matrices with entries in $\{0,1\}$. We say that $\mathfrak{X}$ is an \emph{association scheme} if the following hold:
\begin{enumerate}[(1)]
\item $A_0 = I$, the identity matrix;
\item $\sum_{j=0}^d A_j = J$, the all-ones matrix (In other words, the $1$'s in the $A_j$'s partition $X\times X$);
\item For each $j$, $A_j^T \in \mathfrak{X}$; and
\item For each $i,j$, $A_i A_j \in \spn\mathfrak{X}$.
\end{enumerate}
A \emph{commutative} association scheme also satisfies
\begin{enumerate}[(1)]
\setcounter{enumi}{4}
\item For each $i,j$, $A_i A_j = A_j A_i$.
\end{enumerate}
\begin {example} The Johnson scheme J (v,k). The vertex set of this scheme is the set of all k-subsets of a fixed set of v elements. Two vertices $\alpha$ and $\beta$ are i-related if $\|\alpha \cap \beta\| = k - i$. This scheme has k classes.
\end {example}

\begin {example} \label {ex: Grassmann}
The Grassmann scheme $J_q (v, d)$. The vertex set is the set of all subspaces of dimension d of the vector space of dimension n over GF(q) (finite field with q elements). Subspaces $\alpha$ and $\beta$ are i-related if $dim(\alpha \cap \beta) = i$. This q-deformed Johnson scheme has d classes, may be thought of as a discrete version of a Grassmannian manifold, and the graph it generates is distance transitive and the basis for our construction of an IFS.
\end {example}
Finite groups induce association schemes and the corresponding algebras along with them. For example, let $X=G$ be a finite group, and let $K \subset \Aut(G)$ be a group of automorphisms of $G$. Let $\{e\} = C_0,\dotsc, C_d$ be the orbits of $K$ acting on $G$. If $\{A_x\}_{x\in G}$ define $B_0,\dotsc,B_d$ by
\[ B_j := \sum_{x\in C_j} A_x. \]
Then $\mathfrak{X}:=\{B_j\}_{j=0}^d$ is an association scheme. We call this a \emph{subscheme} of $\{A_x\}_{x\in G}$.
The \emph{adjacency algebra} of an association scheme $\{A_j\}_{j=0}^d$ is $\mathscr{A}:=\spn\{A_j\}_{j=0}^d$ is a \emph{Bose-Mesner algebra}. It is a unital $*$-algebra of matrices and hence a von Neumann algebra. It is also closed under the Hadamard (Schur) product that we can turn into a quantum probability space by defining a suitable linear functional on it.
When $K$ is the group of inner automorphisms of $G$ (i.e.\ conjugations by elements of $G$), the orbits $C_0,\dotsc,C_d$ are precisely the conjugacy classes of $G$. Then $\mathscr{B}:=\spn\{B_0,\dotsc,B_d\}$ is the center of the group von Neumann algebra $\mathscr{A}:=\spn\{A_x : x \in G\}$.

When $\{A_j\}_{j=0}^d$ is a commutative association scheme the matrices $A_0,\dotsc,A_d$ that form the Bose-Mesner algebra $\mathscr{A}$ has an alternative basis $E_0,\dotsc,E_d$ of projections onto the maximal common eigenspaces of $A_0,\dotsc,A_d$. The algebra $\mathscr{A}$ is closed under the Schur (Hadamard) product has \emph{Krein parameters} (fusion rules in our case) $q_{i,j}^k$ satisfying the relation
\[ E_i \circ E_j = \frac{1}{|X|} \sum_{k=0}^d q_{i,j}^k E_k \qquad (0 \leq i,j \leq d). \]
of the association scheme. The dual notion to Krein parameters is the Intersection numbers $p^k_{ij}$ in terms of the usual matrix product
$A_i \bullet A_j= \sum_{k} p^k_{ij}A_k$.
This parameter corresponds to in a distance-regular graph (ex: complete graphs, cycles, and odd graphs) the number of paths between a pair of k-distant vertices via i-distant plus j-distant paths is independent of the pair.

Now, we can relate an anyonic system with association schemes using the example of the Ising model. 

We have the rules for setting up our Bose-Mesner algebra with Schur product and we refer to the elements of the algebra with the same $\sigma, \psi, 1$ notation. The fusion rules of equation \eqref {FusionRules} are the Krein parameters of the algebra that describe the adjacency 3x3 matrix of a family of four possible graphs $\mathscr{A} = \{A_1, A_2, A_3, A_4\}$ along with the unit element. For example, $A_1 = \begin{bmatrix} 1 & 0 & 0 \\ 0 & 0 & 0 \\ 0 & 0 & 0 \end {bmatrix}$ and $A_2 \begin{bmatrix} 0 & 1 & 0 \\ 1 & 0 & 0 \\ 0 & 0 & 0 \end {bmatrix}$ and of course these two are orthogonal under Hadamard multiplication. The polynomials $A^n_i, i = 1, ..., 4$ encode the n-distance paths of the quantum walks of the anyons that is equivalent to cascading pair-of-pants n-times. Our quantum probability space is $(\mathscr{A}, \rho)$ where $\rho$ is a linear map on the algebra $\mathscr{A}$ satisfying some regularity conditions. For example, a state can assign the subgraphs $N^1_{\sigma \sigma} = 1$, $N^{\psi}_{\sigma \sigma} = -1$, and zero for the rest of the basis of the algebra, by linearity the map can be extended to the whole algebra, corresponding to a qubit in quantum information processing \cite {Radbalutq}. If desired, a Hilbert space can be derived from the von Neumann algebra via the GNS construction and the basis of the algebra form the projections of the Hilbert space. For this state, the matrix algebra can be described by 2x2 matrices and the Braid matrices act as automorphisms.

What we have achieved is an algebraic perspective and a graphical picture on anyonic topological quantum computations based on association schemes. This opens up the possibility suitable for asymptotic analysis of the anyon evolutions via quantum probability.
\

\section {Interacting Fock spaces}
Interacting Fock spaces are one type of generalization of the usual symmetric and anti-symmetric Fock spaces that have applications in quantum optics and graph theory. The IFS framework can be used to model bosonic fields with white noise processes in quantum optics. Unlike in the case of classical probability theory there are several different stochastic independence that can be formulated in the quantum context within the framework of IFS. This independence is required to define graph products and based on the monadic operation different stochastic independence arise and correspondingly various central limit theorems manifest. In a quantum probability space $(\mathscr{A}, \phi)$ the usual commutative independence $(\phi(bab) = \phi(a)\phi(b^2); a,b \in \mathscr{A})$ such as the one assumed in quantum optics leads to conjugate Brownian motions (measured as quadratures) in the limit. The monotone independence $(\phi(bab) = \phi(a)\phi(b)^2; a,b \in \mathscr{A})$ that is relevant in quantum walks leads to arcsin-Brownian motion (double-horn distribution) aymptotically and the other two are free and Boolean independences not focused in this work. In the graph context, the independence notions are defined in terms of products of graphs.
\begin {defn} \cite {Obata2007} An IFS associated with the Jacobi sequences $\{\omega_n\}, (\omega_m = 0) \Rightarrow \forall n \ge m, \omega_n = 0, \{\alpha_n\}, \alpha_n \in \mathbb{R}$ is a tuple $(\Gamma \subset \mathscr{H}, \{\Phi_n\}, B^+, B^-, B^\circ)$ where $\{\Phi_n\}$ are orthogonal polynomials and $B^\pm \Phi_n$ spans $\Gamma$, the subspace of the Hilbert space $\mathscr{H}$. The mutually adjoint operators $B^+, B^-$ and $B^\circ$ satisfy the relations
\begin {align*}
B^+ \Phi_n = \sqrt{\omega_{n+1}} \Phi_{n+1}. \\
B^- \Phi_n = \sqrt{\omega_n} \Phi_{n-1}; B^- \Phi_0 = 0. \\
B^\circ \Phi_n = \phi_n.
\end {align*}
\begin {equation} 
xP_n (x) = P_{n+1} (x) + \omega_n P_{n-1} (x) + \alpha_{n+1} P_n (x).
\end {equation}
\end {defn}
With the above IFS we can associate a graph with an adjacency matrix 
$T = \begin{bmatrix} \alpha_1 & \sqrt{\omega_1}  \\
\sqrt{\omega_1} & \alpha_2 & \sqrt{\omega_2}  \\
& \sqrt{\omega_2} & \alpha_3 & \sqrt{\omega_3}  \\
 & & \ddots & \ddots & \ddots & \\
& & & \sqrt{\omega_{n-1}} & \alpha_n & \sqrt{\omega_n} \\
  & & & & \ddots & \ddots & \ddots 
\end {bmatrix}$
that has the quantum decomposition $T = B^+ + B^- + B^\circ$. The sequence $\{\Phi_n\}$ represents fixing a vertex and stratifying (partitioning based on distance from the fixed vertex) the graph with V set of vertices. Let us fix the Hilbert space $\mathscr{H} = l^2(V)$ of the graph for the rest of the section. The above Jacobi matrix has the structure of next-neighbor hopping Markov chain and this fact will be used later to make the connection to subfactors. 

\begin{figure}
 \center {
 \begin{tikzpicture}[->,>=stealth',shorten >=1pt,auto,node distance=1.5cm,
  thick,main node/.style={ellipse,fill=green!20,draw,font=\sffamily\Large\bfseries}]
  \node[main node] (1) {\hspace{0.5cm} };
  \node[main node] (2) [right of=1] {\hspace{0.5cm } };
  \node[main node] (3) [below of=1] {0};
  \node[main node] (4) [below of=2] {1};
  \node[main node] (5) [right of=2] {\hspace{0.5cm } };
  \node[main node] (6) [below of=5] {2};
  \node[main node] (7) [right of=5] {\hspace{0.5cm } };
  \node[main node] (8) [below of=7] {3};
  \node[main node] (9) [right of=7] {};
  \node[main node] (10) [below of=9] {};
  \node[main node] (11) [right of=9] {\hspace{0.5cm } };
  \node[main node] (12) [below of=11] {\hspace{0.5cm } };
  \node[main node] (13) [right of=11] {\hspace{0.5cm } };
  \node[main node] (14) [below of=13] {m};
  \node[main node] (15) [right of=13] {\hspace{0.5cm } };
  \node[main node] (16) [below of=15] {\hspace{0.5cm } };

  \path[every node/.style={font=\sffamily\small}]
    (1) edge[-] node {}  (2)    
    (2) edge[-] node {} (4)        
          edge[right] node[right=1mm] {}(6)
    (3) edge[-] node[right=1mm] {}(1)
          edge[left] node[right=1mm] {}(2)        
    (4) edge[-] node[right=1mm] {}(3)
         edge[-] node[right=1mm] {} (6)
    (5) edge[-] node[right=1mm] {} (6)
         edge[-] node[right=1mm] {} (2)
    (7) edge[-] node[right=1mm]{} (8)
         edge[-] node[right=1mm] {}(5)
    (6) edge[-] node[right=1mm] {}(8)
          edge[right] node[right=1mm] {}(7)
    (8) edge[-] node[right=1mm]{} (10)
    (7) edge[-] node[right=1mm]{} (9)
          edge[right] node[right=1mm]{} (12)
    (9) edge[-] node[right=1mm]{} (11)
    (11) edge[-] node[right=1mm]{} (12)
           edge[-] node[right=1mm]{} (13)
    (10) edge[-] node[right=1mm]{} (12)
    (12) edge[-] node[right=1mm]{} (14)
            edge[right] node[right=1mm]{} (13)
    (13) edge[-] node[right=1mm]{} (14)
            edge[right] node[right=1mm]{} (16)
            edge[-] node[right=1mm]{} (11)
    (15) edge[-] node[right=1mm]{} (16)
             edge[-] node[right=1mm]{} (13)
    (14) edge[-] node[right=1mm]{} (16);
    \end{tikzpicture}
  
  \caption{\label{fig:CoinGraph}
     The m-moment of Bernoulli trial represented as a random walk on a graph as an m-step evolution starting and ending at the same node. The directed part may be viewed as a comb product of two graphs each with two nodes and the whole graph may be seen as cartesian product of the two graphs. In the comb product case (monotone stochastic independence), quantum walk evolution, the asymptotic distribution is arcsin-Brownian motion and in the cartesian product case (commutative independence) the long time limit is a non-commutative Brownian motion. 
  }
}
\end{figure}
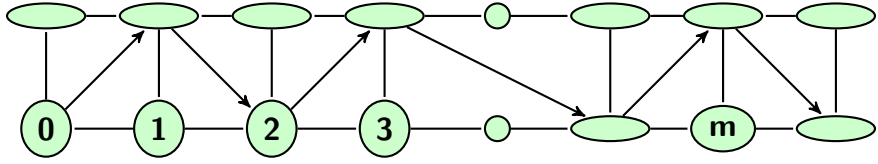

\begin {figure}
\begin{tikzpicture}[->,>=stealth',shorten >=1pt,auto,node distance=2cm,
  thick,main node/.style={circle,fill=blue!20,draw, 
  font=\sffamily\Large\bfseries,minimum size=10mm},
  sec node/.style={draw}]
  \node[main node] (0) {0};
  \node[main node] (1) [right of=0] {1};
  \node[main node] (2) [right of=1] {2};
  \node[sec node] (3) [right of=2] {...};
  \node[main node] (4) [right of=3] {n};
  \node[sec node] (5) [right of=4] {...};
  \path[every node/.style={font=\sffamily\small,
  		fill=white,inner sep=1pt}]
    (0)  edge [loop below] node {$b_0$} (0)
          edge [bend left=60] node[right=1mm] {$c_1$} (1)
    (1)  edge [loop below] node {$b_1$} (1)
          edge [bend left=50] node[left=1mm] {$a_0$} (0)
          edge [bend left=60] node[right=1mm] {$c_2$} (2)
    (2) edge [loop below] node {$b_2$} (2)
          edge [bend left=50] node[left=1mm] {$a_1$} (1)
          edge [bend left=60] node[right=1mm] {$c_2$} (3)
    (3) edge [bend left=50] node[left=1mm] {$a_2$} (2)
         edge [bend left=60] node[right=1mm] {$c_n$} (4)
    (4) edge [loop below] node {$b_n$} (4)
          edge [bend left=50] node[left=1mm] {$a_n$} (3)
          edge [bend left=50] node[right=1mm] {} (5)
    (5) edge [bend left=50] node[left=1mm] {} (4);
 \end{tikzpicture} 
\caption{\label{fig:digraph} Weighted digraph with an adjacency matrix } 
$T = \begin{bmatrix} b_0 & a_0 & & & &  & \\
c_1 & b_1 & a_1 & & &  & \\
 & \ddots & \ddots & \ddots & \\
  & & c_n & b_n & a_n & \\
  & & & \ddots & \ddots & \ddots 
\end {bmatrix}$
\end {figure}
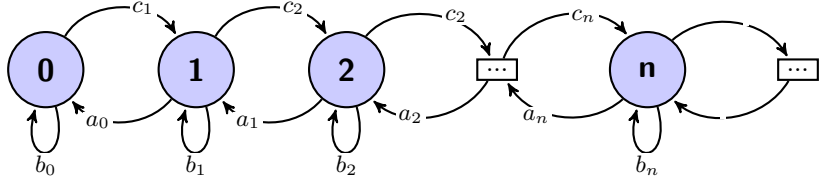

\begin {example} \label {ex: Bern} Bernoulli trial (Figure \ref {fig:CoinGraph}): This process produces an ensemble that is a classical coin toss with the probability measure $\mu = \frac{1}{2}\delta_{-1} + \frac{1}{2}\delta_1$ whose moment sequence is $M^m_\mu = \int_\infty ^\infty x^m \mu(x) = 1$ if m is even and 0 when m is odd. Now, let us consider a graph with two nodes $(e_0, e_1)$ with an edge connecting them. The adjacency matrix for the graph in the standard basis ($\{\begin{bmatrix} 0 \\ 1 \end{bmatrix}, \begin{bmatrix} 1 \\ 0 \end{bmatrix}\}$) of $\mathbb{C}^2$ is $T =  \begin{bmatrix} 0 & 1 \\ 1 & 0 \end{bmatrix}$. In quantum probability we have $\langle e_0, A^m e_0 \rangle = $ if m is even and 0 when m is odd, that is, it reproduces the classical probability measure. In other words, the quantum random variable A reproduces in the vacuum state $e_0$ the moment sequence of the classical coin toss and the fact that it has the canonical decomposition $A = \begin{bmatrix} 0 & 1 \\ 0 & 0 \end{bmatrix} + \begin{bmatrix} 0 & 0 \\ 1 & 0 \end{bmatrix}$ we can say that Bernoulli trial has a quantum decomposition. The classical moment sequence has the new interpretation in the graph context as the number of m-step walks starting from the vertex $e_0$ and ending in it (polynomials of A). The same analysis holds for the state $e_1$ as the vertices of a graph are equivalent any arbitrary vertex can be used to define the vacuum state. Also when a biased coin, produces heads with probability p and tails with probability (1 - p), is used at the quantum state $\phi(A) = p*\alpha e_0 + (1 - p)*\beta e_1$, where $A = \alpha \ket{e_0}\bra{e_0} + \beta \ket{e_1}\bra{e_1}, \alpha^2 + \beta^2 = 1$ the same ensemble is produced. In Hadamard quantum walk on the integer line the observation that the probability amplitudes vanish at odd positions can be understood from the non-existent odd-moments interpretation. So, growing a distance regular graph preserves the moments, of all possible order, information and taking the limit on the size of the graph would provide the asymptotics distribution. 
\end {example}
\begin {example} A weighted directed graph (Figure \ref {fig:digraph}) that is an infinite Markov chain can be represented by an adjacency matrix and an IFS.
\end {example}

\begin {example}
For the bosonic (symmetric) Fock space we have $\omega_n = n; \alpha_n = 0$.
For the fermionic (anti-symmetric) Fock space the Jacobi parameters are $\omega_1 = 1; \omega_n =0, n > 1; \alpha_n = 0$.
\end {example}

In this section we construct an IFS for the association scheme with d classes $\mathfrak{X}:=\{B_j\}_{j=0}^d$ similar to the development of multi-dimensional orthogonal polynomials by Accardi \cite{Accardi2017}. For example, association schemes induced by finite cyclic groups are commutative and as a consequence the schemes are self-duals. In our case the variables are matrices and the algebra is also closed under Schur multiplication $\circ$ and thus a *-algebra with the ladder operators (CAPs) of the IFS can be defined in terms of the parameters of the association scheme. These are positive real numbers and can be normalized to become a probability measure with their square root interpreted as probability amplitudes. We will work with the induced hypergoup of the conjugacy classes where the Krein numbers can be interpreted as collision probabilities. The classes of the association schemes are referred to as modes and they represent different graphs with common vertices. Orthogonal polynomials in finite number of variables were treated in \cite {Stan2004} and the commutation relations between the ladder operators derived and here our focus is determine the Jacobi parameters. IFS are the natural framework to define quantum walks on graphs and then to establish their aysmptotics \cite {Konno2013} \cite{Obata2007}.

\newcommand{\D}{7} 
\newcommand{\U}{7} 

\newdimen\R 
\R=3.5cm 
\newdimen\L 
\L=4cm

\newcommand{\A}{360/\D} 

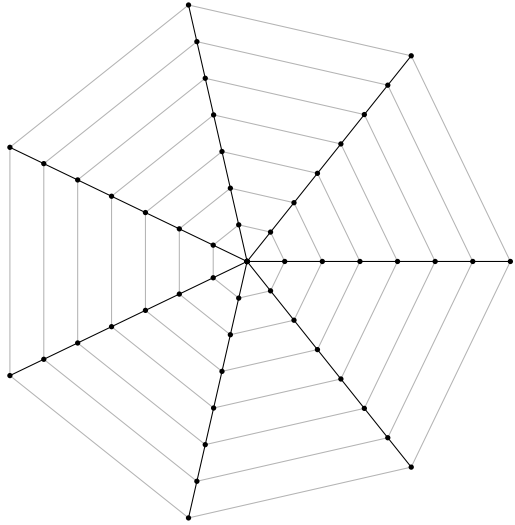
\begin{figure}[htbp]
 \centering

\begin{tikzpicture}[scale=1]
  \path (0:0cm) coordinate (O); 

  \foreach \X in {1,...,\D}{
    \draw (\X*\A:0) -- (\X*\A:\R);
  }

  \foreach \Y in {0,...,\U}{
    \foreach \X in {1,...,\D}{
      \path (\X*\A:\Y*\R/\U) coordinate (D\X-\Y);
      \fill (D\X-\Y) circle (1pt);
    }
    \draw [opacity=0.3] (0:\Y*\R/\U) \foreach \X in {1,...,\D}{
        -- (\X*\A:\Y*\R/\U)
    } -- cycle;
  }

  \path (1*\A:\L) node (L1) {};
  \path (2*\A:\L) node (L2) {};
  \path (3*\A:\L) node (L3) {};
  \path (4*\A:\L) node (L4) {};
  \path (5*\A:\L) node (L5) {};
  \path (6*\A:\L) node (L6) {};
  \path (7*\A:\L) node (L7) {};

  %
  %

  %

\end{tikzpicture}
\caption{Spiderweb Diagram - an example of a stratified graph on which a quantum walk can be fashioned.}
\label{fig:spiderweb}
\end{figure}

Let us denote the *-algebra of matrices on $B_i\in\mathfrak{X}$ as $\mathcal{P} = \mathbb{C}[(B_i),0\leq{n}\leq{d}]$ which are Schur polynomials.  The linear generators of the algebra are the monomials $M = B^{n_1}_1,\dots,B^{n_d}_d$ and $deg(M) = \sum_j (n_j) = n$ with $1_\mathcal{P}$ as as unity satisfying $\langle{1}_\mathcal{P},{1}_\mathcal{P}\rangle = 1$.  We get the symmetric tensor commutative algebra (bosonic) $\sum_{n\in\mathbb{N}}\mathcal{P}_n^0$ that is graded by the correspondences $e_j\in\mathbb{C}^d\rightarrow{B_j}$ and $\otimes_{sym}(\mathbb{C}^d) = \sum_{n\in\mathbb{N}}\mathcal{P}_n^0 \equiv\mathcal{P}$. Here, $\mathcal{P}_n^0$ is the span of monomials of degree n. We can also build a $\mathbb{Z}_2$-graded *-algebra using anti-symmetric tensor products (fermionic) as $\otimes_{asym}(\mathbb{C}^d) = \sum_{n\in\mathbb{N}}\mathcal{P}_n^0 \equiv\mathcal{P}^a$. The natural pre-inner product $\langle{B_1},B_2\rangle = tr(B_1^*{B_2}) = sum(B_1\circ{B_2})$ of our Bose-Mesner algebra extends to a pre-inner product on gradations of $\mathcal{P}$ which in turn induces a state $\phi(B_1^*B_2) = tr(B_1^*{B_2})$. Gradation is an algebraic property independent of the measure but when orthogonality is based on the state $\phi$, $\mathcal{P}$ produces orthogonal quantum decomposition of $B_i$s. It is obviously a product state on $\mathcal{P}$ and the filtration is constructed in the usual way. 
\begin {align*}
\mathcal{P}_{n]} &= \text {linear span of monomials of degree n}. \\
P_{n]}:&\mathcal{P}\rightarrow \mathcal{P}_{n]}. \\
P_n &= P_{n]} - P_{n - 1]}.
\end {align*}
We have,
\begin {align*}
a^+_{j\mid{n}} &= P_{n+1} X_j P_n. \\
a^-_{j\mid{n}} &= P_{n-1} X_j P_n. \\
a^+0_{j\mid{n}} &= P_{n} X_j P_n. \\
a^\epsilon_j &= \sum_j a^\epsilon_{j\mid{n}}. \\
B_j &= a^+_j + a^0_j + a^-_j. \\
tr(B_j^*, B_i) &= 0.
\end {align*}

The Jacobi relation in this case is similar to the polynomials in real indeterminates and given by:
\begin {align} \label {eq:Jacobi-3-term}
B_j P_n &= P_{n+1}B_j P_n + P_n B_j P_n + P_{n-1} B_j P_n; 1\leq{j}\leq{d}. \\
B_j P_n &= a^+_{j\mid{n}} + a^0_{j\mid{n}} + a^-_{j\mid{n}}. \\
\end {align}
Our IFS, with generalized Jacobi parameters as positive definite kernels and hermitian matrices, is fashioned on the procedure described in \cite{Accardi2017b} where the n-th level spaces defined recursively in terms of (n-1)-th level spaces. Let us the define the pre-Hilbert space that is a linear span of product vectors of individual modes $\{(\Phi_{1,n}.\Phi_{2,n}\dots\Phi_{d,n}.\Phi_0)\}$ with the vacuum vector:
\begin {align*}
\Phi_0 &= 1_\mathcal{P}. \\
\Phi_{j,0} &= \Phi_0, 0\leq{j}\leq{d}.
\end {align*}

\begin {align*}
\Phi_{j,n} &=  a^+_{jn}\dots{a}^+_{j1}\Phi_0. \\
a^0_{j\mid{0}}\Phi_0 &= p^0_{1j}\Phi_0.\\
a^+_j\Phi_{j,n} &= \sqrt{p^{n+1}_{j;1,n}p^{n}_{j;1,n+1}}\Phi_{j,n+1}.\\
a^-_j\Phi_{j,n} &= \sqrt{p^{n}_{j;1,n-1}p^{n-1}_{j;1,n}}\Phi_{j,n-1}; a^-\Phi_{j,0} = 0.\\
a^o_j\Phi_{j,n} &= p^n_{j;1,n}\Phi_{j;1,n}.\\
\end {align*}
We have d CAPs (creation, annihilation, and preservation operators) one for each mode of the system and they are related to the parameters as follows:
In the case of 1-D polynomials the stratification of the corresponding graph, creation, annihilation, and preservation operators of the algebra can be defined on $\mathscr{A}$ using the natural ordering of the association scheme $0\leq{n}\leq{d}$ as
\begin {align*}
V_n &= \{x\in{C_i}\}.\\
\Phi_n &= \|V_n\|^{-\frac{1}{2}}\sum_{x\in{V_n}}\delta_x.\\
\omega_\epsilon(x) &= p^n_{j;1,n+\epsilon}, \text {  if  } x\in{V_n}. 
\end {align*}
 We have the following result on constructing an IFS asymptotically on a growing graph:

\begin {theorem}
Let $\mathscr{A}$ be a Bose-Mesner algebra corresponding to the association scheme $\{A_i\}_{1\leq{i}\leq{d}}$ with the intersection numbers $\{p^k_{ij}\}$ generated by a group $\mathscr{G}$ and $\Gamma(\mathscr{G}) = span(\Phi_d)$. Then, $\Gamma(\mathscr{G})$ is asymptotically invariant under the action of $A^\epsilon=a^\epsilon_1{a^\epsilon_2}\dots{a^\epsilon_d}, \epsilon\in\{+,-,o\}$ and $(\Gamma(\mathscr{G}), \{|phi_n\}, A^+, A^-)$ is an interacting Fock space.
\end {theorem}
%

We can set up a Grover quantum walk on the IFS constructed out of a growing graph $\mathscr{A}$ as follows:

\begin {align*}
G &= (V, E);  &\text{ regular graph }. \\
A(G) &= \{(x, y) \in V \times V; x~y\}. &\text{ set of half-edges}. \\
\mathscr{H} &= l^2(A(G)); &\text {state space of Grover walk}. \\
\{ \widetilde{\delta}_{(u, v)}; &(u, v) \in A(G) \}; &\text {canonical basis of } \mathscr{H}. \\
C\widetilde{\delta}_{(u, v)} &= \sum_{w:(u,v)\in A(G)} (H^{(u)})_{uv}\widetilde{\delta}_{(u, v)}; &\text {Coin flip operator}.\\
(H^{(u)})_{uv} &= \frac{2}{deg(u)} - \delta_{wv}; &\text { Grover matrix}.\\
S\widetilde{\delta}_{(u, v)} &= \widetilde{\delta}_{(v, u)}; &\text { Shift operator}. \\
U &= SC; & \text { unitary evolution for the quantum walk}. \\
&\{ \Phi_n = U^n \Phi_0 \}; & \text { Grover walk on IFS}.
\end {align*}
The above construction of an IFS would provide a growing graph with probability measures defined on the adjacency matrix algebra. What we need is a way to incorporate probability amplitudes while we grow the graph via anyonic fusions. The anyonic system to generate a regular graph can be constructed as follows: We consider the dual to the intersection numbers $\{p_{ij}^k\}$
of the Bose-Mesner algebra $\mathscr{A}$ which are the Krein parameters $\{q_{ij}^k\}$ that provide the fusion rules of the anyonic system.
That is, the dimensions of the Hilbert spaces to define the pants axiom of the anyon system are nothing but the Krein numbers that can be derived from the dual intersection numbers. When the anyons are allowed to interact randomly they will eventually build the distance-regular graph. What we need is a unitary quantum walk with this anyonic system satisfying the above Grover evolution of probability. 

To illustrate this Grover walk on a simpler graph let us consider a homogeneous tree of degree three that is an infinite graph and fashion a quantum walk on it. This infinite graph (Figure \ref {fig:QW-1}) is stratified as an IFS with the vacuum state $\Phi_0$ (first stratum) having a Dirac mass at the origin. The intersection numbers for the system is given by:
\begin {align*}
p_{11}^0 &= 3 &. \\
p_{1,n}^{n + 1} &= 2. &n = 2,3, \dots, \\
p_{1,n}^{n - 1} &= 1. &n = 2,3, \dots, \\
p_{1,n}^n &= 0. &n = 2,3, \dots, \\
\end {align*}
The corresponding krein parameters will be non zero only for the above cases and so from each anyon corresponding to a strata there is only one way to go to the next stage. As we grow the graph using anyons in the dual picture we are not encoding any probability amplitudes. We can have $\sigma$ anyons on fixed locations of the hexagonal lattice and grow the graph and then incorporate probabiity amplitudes corresponding to the quantum walk evolution with additional fusions and braidings.
Each strata of the graph could be associated with an association scheme class that will lead to infinite number of them and the corresponding number of topological charges (anyons) would be infinite. We can exploit the symmetry in the system to create topological charges corresponding to successive strata by fusion rules of simpler anyons. This may not be possible for an arbitrary graph but for distance regular ones. We can use a system with two anyons and encode the graph structure in the fusion space of them. 
The second and third intersection numbers above, that connects two adjacent strata of a regular graph or two anyonic topological charges in the dual picture,  provide a clue that there are two non trivial fusion rules. The corresponding Bose-Mesner algebra has three generators and the dual anyonic system with three types is given by $(1, \sigma, \psi)$ with $S = \begin{bmatrix} 1 & \sqrt{2} & 1 \\ \sqrt{2} & 0 & -\sqrt{2} \\ 1 & -\sqrt{2} & 1 \end {bmatrix} $, $\theta_\sigma = e^{i\pi/8}$, $\theta_\psi =  -1$. This system of anyons has the following fusion rules:
\begin {align*}
\sigma \times \sigma &= 1 + \psi. \\
\sigma \times \psi &= \sigma. \\
\psi \times \psi &= 1. \\
\psi \times 1 &= \psi. \\
\sigma \times 1 &= \sigma.
\end {align*}
We need to fuse anyons in a such a way to have the same target topological charge, by any combination of fusion outcomes, so that superposition is possible to create mixed states. Three $\sigma$ anyons are required to encode a qubit as seen below:
\begin {align*}
\sigma \times \sigma \times \sigma = (1 + \psi)\sigma &= 2\sigma. \\
\sigma \times \sigma \times \sigma \times \sigma &= 2(1 + \psi). \\
\sigma \times \sigma \times \sigma \times \sigma \times \sigma &= 4\sigma.
\end {align*}
So we need to fuse three $\sigma$ anyons to encode a qubit and six of them to represent a three state qutrit. 

The quantum walk with the Hilbert space $\mathscr{H} = \mathbb{C}^3\otimes l^2(A(G))$ shown in the following three figures (first three steps) propagate quantum probabilities to outer strata. the number of nodes of strata form the sequence (1, 3, 6, 12, 24, ...) and we need fixed $\sigma$ anyons at each of these nodes. We can create them at each position by fusion of three $\sigma$ anyons. In addition to represent the qutrit coin we need six $\sigma$ anyons for each node with following encoding:
\begin {align*}
\ket{0} &= \ket{\sigma\sigma\rightarrow 1}\ket{\sigma\sigma\rightarrow 1}\ket{\sigma\sigma\rightarrow 1}.\\
\ket{1} &= \ket{\sigma\sigma\rightarrow 1}\ket{\sigma\sigma\rightarrow \psi}\ket{\sigma\sigma\rightarrow \psi}.\\
\ket{2} &= \ket{\sigma\sigma\rightarrow \psi}\ket{\sigma\sigma\rightarrow \psi}\ket{\sigma\sigma\rightarrow 1}.
\end {align*}
In priciple six anyons can encode four states but we need only three. We can use braiding for the coin operation to evolve the topological quantum walker. The walk will start with vacuum splitting into an anyon and its anti-particle, that is a graph with a single edge (Figure \ref {fig:QW-1}). In the second step three anyons will fuse the first anyon with respective probability amplitudes (2/3, 2/3, -1/3) (Figure \ref {fig:QW-2}) and increase the graph depth to two. In the third step, similar fusion steps will generate a regular graph of depth three (Figure \ref {fig:QW-3}). And the process repeats infinitely generating the probability amplitudes on a growing graph. Such an evolution can be realized using Ising systems \cite {Ebadi2021} that supports braiding and the required (for more general coin than Grover) probability amplitudes can be achieved by a combination of braiding and topological spins in addition to fusion.
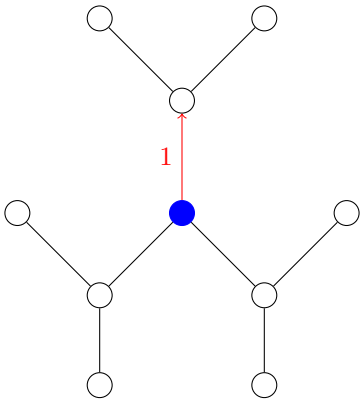
\begin{figure}[htbp]
 \centering
\begin{tikzpicture}[node distance =0.85 cm and 0.85 cm]
    \node[draw,circle,blue,fill] (1) [below] {};
    \node[draw,circle] (2) [above = of 1,yshift=0.3cm] {};
    \node[draw,circle] (3) [below left = of 1] {};
    \node[draw,circle] (4) [below right = of 1] {};
    \node[draw,circle] (5) [above right = of 2] {};
    \node[draw,circle] (6) [above left = of 2] {};
    \node[draw,circle] (7) [below  = of 3] {};
    \node[draw,circle] (8) [above left = of 3] {};
    \node[draw,circle] (9) [below  = of 4] {};
    \node[draw,circle] (10) [above right = of 4] {};
    \path (2) edge node[above] {} (5);
    \path (2) edge node[above  ] {} (6);
    \draw[->,red] (1) edge node[auto] {\(1\)} (2);
    \path (1) edge node[above  ] {} (3);
    \path (1) edge node[above  ] {} (4);
    \path (8) edge node[above  ] {} (3);
    \path (7) edge node[above  ] {} (3);
    \path (10) edge node[above  ] {} (4);
    \path (9) edge node[above  ] {} (4);

\end{tikzpicture}
\caption{Step one of quantum walk with the edges labeled by the respective probability amplitudes.}
\label{fig:QW-1}
\end{figure}
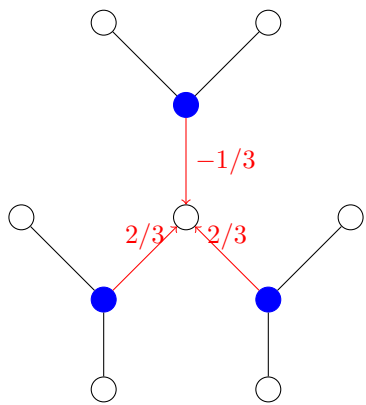
\begin{figure}[htbp]
 \centering
\begin{tikzpicture}[node distance =0.85 cm and 0.85 cm]
    \node[draw,circle] (1) [label = below: ] {};
    \node[draw,circle,blue,fill] (2) [above = of 1,yshift=0.3cm] {};
    \node[draw,circle,blue,fill] (3) [   below left = of 1 ] {};
    \node[draw,circle,blue,fill] (4) [   below right = of 1] {};
    \node[draw,circle] (5) [   above right = of 2] {};
    \node[draw,circle] (6) [   above left = of 2] {};
    \node[draw,circle] (7) [   below  = of 3] {};
    \node[draw,circle] (8) [   above left = of 3] {};
    \node[draw,circle] (9) [   below  = of 4] {};
    \node[draw,circle] (10) [   above right = of 4] {};
    \path (2) edge node[above  ] {} (5);
    \path (2) edge node[above  ] {} (6);
    \draw[->,red] (2) edge node[auto] {\(-1/3\)} (1);
    \path[->,red] (3) edge node[above  ] {\(2/3\)} (1);
    \path[->,red] (4) edge node[above  ] {2/3} (1);
    \path (8) edge node[above  ] {} (3);
    \path (7) edge node[above  ] {} (3);
    \path (10) edge node[above  ] {} (4);
    \path (9) edge node[above  ] {} (4);

\end{tikzpicture}
\caption{Step two of quantum walk with the edges labeled by the respective probability amplitudes.}
\label{fig:QW-2}
\end{figure}
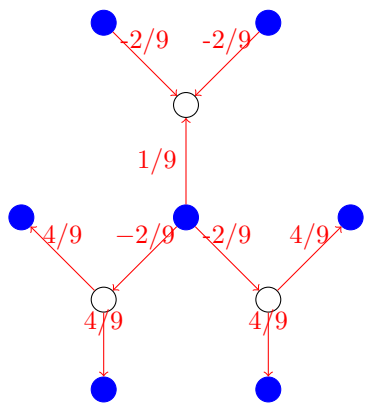
\begin{figure}[htbp]
 \centering
\begin{tikzpicture}[node distance =0.85 cm and 0.85 cm]
    \node[draw,circle,blue,fill] (1) [] {};
    \node[draw,circle] (2) [above = of 1,yshift=0.3cm] {};
    \node[draw,circle] (3) [   below left = of 1 ] {};
    \node[draw,circle] (4) [   below right = of 1] {};
    \node[draw,circle,blue,fill] (5) [   above right = of 2] {};
    \node[draw,circle,blue,fill] (6) [   above left = of 2] {};
    \node[draw,circle,blue,fill] (7) [   below  = of 3] {};
    \node[draw,circle,blue,fill] (8) [   above left = of 3] {};
    \node[draw,circle,blue,fill] (9,red,fill) [   below  = of 4] {};
    \node[draw,circle,blue,fill] (10) [   above right = of 4] {};
    \path[->,red] (5) edge node[above  ] {-2/9} (2);
    \path[->,red] (6) edge node[above  ] {-2/9} (2);
    \draw[->,red] (1) edge node[auto] {\(1/9\)} (2);
    \path[->,red] (1) edge node[above  ] {\(-2/9\)} (3);
    \path[->,red] (1) edge node[above  ] {-2/9} (4);
    \path[->,red] (3) edge node[above  ] {4/9} (8);
    \path[->,red] (3) edge node[above  ] {4/9} (7);
    \path[->,red] (4) edge node[above  ] {4/9} (10);
    \path[->,red] (4) edge node[above  ] {4/9} (9);

\end{tikzpicture}
\caption{Step three of the Grover quantum walk with the edges labeled by the respective probability amplitudes.}
\label{fig:QW-3}
\end{figure}
\

The connection between distance-regular graphs and anyon systems can be understood by looking at the intersection numbers of the IFS. The strata of a distance-regular graph are orthogonal projections of the IFS and the intersection numbers indicate that hopping is only possible between nearest neighbors. This structure is the same as that of projections of Temperley-Lieb algebras that are invariants of subfactors. In the next step we will unveil this structure of distance-regular graphs and cast it in the language of subfactor theory and establish that the connection between such graphs and anyon systems is a natural one.

\section {Summary and Conclusions}
We investigated distance regular graphs using association schemes that are endowed with a duality that describes particle collision in one picture and growing graph in another. We constructed IFS on growing graphs with possible quantum walk evolutions and identified anyon based realizations exploiting the duality. In the next step we plan to establish the connection using Temperley algebras and subfactors in a more rigorous setting.
\

\bibliographystyle{abbrv}

\end{document}